\documentclass[%
 reprint,
 amsmath,amssymb,
 aps,
]{revtex4-1}
\usepackage{graphicx}
\usepackage{multirow}
\usepackage{dcolumn}
\usepackage{bm}
\usepackage{latexsym}
\usepackage{wasysym}
\usepackage{amsmath}
\usepackage{MnSymbol}
\usepackage{natbib}
\usepackage{tabularx}

\begin{document}

\preprint{APS/123-QED}

\title{Blending of Conceptual Physics and Mathematical Signs}
\thanks{Portions of this paper previously appeared in preliminary work \cite{huynh2018blending}.}%

\author{Tra Huynh}
 \altaffiliation{trahuynh@ksu.edu}
\author{Eleanor C. Sayre}%
 \email{esayre@ksu.edu}
\affiliation{%
 Kansas State University, Physics Department, 116 Cardwell Hall, 1288 N. 17th St. Manhattan, KS 66506-2601 
}%

\date{\today}

\begin{abstract}
Mathematics is the language of science. Fluent and productive use of mathematics requires one to understand the meaning embodied in mathematical symbols, operators, syntax, etc., which can be a difficult task. For instance, in algebraic symbolization, the negative and positive signs carry multiple meanings depending on contexts. In the context of electromagnetism, we use conceptual blending theory to demonstrate that different physical meanings, such as directionality and location, could associate to the positive and negative signs. With these blends, we analyze the struggles of upper-division students as they work with an introductory level problem where the students must employ multiple signs with different meanings in one mathematical expression. We attribute their struggles to the complexity of choosing blends with an appropriate meaning for each sign, which gives us insight into students' algebraic thinking and reasoning. 

\end{abstract}

\maketitle



\section{\label{sec:introduction}Introduction}

Mathematics is important in understanding physics, but many students struggle with integrating mathematical formalism.  The language of mathematics, like other languages, includes homographs and homonyms: a symbol in mathematics can bear multiple meanings, possibly causing trouble for language learners. This can be especially prevalent (or problematic) when mathematics is used in other scientific contexts, such as physics, when the meaning of a symbol can be more specific, yet diverge from the standard mathematical conventions at the same time.

Signs, both negative and positive, are important in both math and physics. Building on Vlassis' work in elementary algebra \cite{vlassis2004making}, a map of how the negative sign is used includes three natures: unary (structural signifier), binary (operating signifier), and symmetric functions. For example, the signs in the metaphor of floors in a building can play different roles, such as moving from one floor to another (binary - `going up two') and establishing the location of a floor (unary - `the second floor'). Successful sense-making of a mathematical expression requires students to flexibly and suitably ascribe meaning to signs.  Yet, algebra students often fail to consider explicitly that the minus sign could have a double status or meaning \cite{vlassis2004making}. 

In physics, signs can signify a direction (``towards''), a flavor (``negative charge''), a location (``$-3\hat{x}$''), or an operation between two terms (``$-kx - cv$'').  The complexity of reasoning with signs in physics, especially negative signs and negative quantities, has focused primarily on the results that reasoning with signs is hard, especially at the introductory level.  However, mechanisms for why reasoning with signs is hard have been elusive.  Moreover, there is both a lack of connection with mathematics literature and a lack of a strong theoretical framework that accounts for a homographic property of signs, which would give us better insights to students' reasoning and struggles.

We pull out two conceptual spaces which might be affiliated with sign in physics: location and direction.  Using the machinery of conceptual blending, we show how those two spaces may blend with sign to produce five different blends. We illustrate the blends in a one-dimensional problem from electrostatics, first with a principled solution and then with two student solutions.  Our goal is to present substantial theoretical discussion grounded in a simple physical scenario, to show how even a simple problem and the relatively limited space of signs have rich conceptual complexity undergirding them. This is not an exhaustive accounting of all that students can do with sign, or even with this problem.

Our work is in contrast to prior work which suggested that two input spaces have only one resulting blend, and that difficulty in blending arises from difficulty choosing appropriate input spaces or running a blend. 

\section{Signs in education research}

Historically, the concept of negativity was attached to the operation of subtraction. In mathematics history, the minus sign is first introduced to describe a quantity yet to be subtracted. A precise difference between the minus sign that assigns a negative number and an operation (subtraction) was not clearly defined until the development of algebra, when negative numbers gained the status of mathematic objects \cite{kilhamn2011making}.

Significant research on how we make sense of negative numbers has been conducted within the context of mathematics education since the 1980s. \citet{lakoff2000mathematics} suggest that negative number sense could be acquired by extending and stretching their set of grounding metaphors for counting-number arithmetic to the set of integers. 
Various instructional models have been developed that employ metaphor to better teach negative numbers and the operations on them to children \cite{kilhamn2011making, ball1993eye}. In contrast, \citet{vlassis2004making} argues that it is not the nature of the negative number itself but the negative sign itself that causes difficulties. 

In physics, negative and positive signs embody additional characteristics for quantities. The meaning of such signs varies specifically with context. 
For example, negativity in reasoning about energy can be particularly challenging \cite{stephanik2012examining, lindsey2014student}. Two common ontologies for energy are substance (energy contained in objects) and vertical location (higher or lower energy level) \cite{scherr2012intuitive, dreyfus2015applying}. Each ontology has both advantages and drawbacks in making sense of different aspects of negativity. For example, the substance ontology is helpful in thinking of the transference of energy but appears to be a hindrance in the case of negative substance or negative energy, where the vertical location ontology is more appropriate and convenient\cite{Dreyfus2012,dreyfus2015applying}. Therefore, accounting for the sign contexts while flexibly choosing the signs' meaning is significant for making sense in physics.  

As a scalar, energy itself does not have a direction.  However, when it comes to vector quantities, positive and negative signs carry the additional meaning of directionality, which has been shown to cause additional trouble for students. Middle school physics teachers inappropriately interpret the signs of acceleration using the speed model \cite{Tabachnick2018}: accelerating is always in the positive direction and decelerating is always in the negative direction.  The speed model makes sense from a person-centric motion perspective: we usually move forwards and think of that as the positive direction.  Walking backwards is rare; if we need to go back, we usually turn around and walk forwards.  

Reasoning about sign for both scalar and vector quantities is complicated by mathematical formalism in physics; in one-dimensional problems, we often treat quantities flexibly as either vectors or scalars.  For example, consider the kinematics equation for final velocity as a function of initial velocity, acceleration, and time:
\begin{align}
\vec{v_f} = \vec{v_i}+\vec{a}t
\end{align}
If an object is slowing down, there must be something subtracted from $\vec{v_i}$ \footnote{In this paper, we use the typographic convention that vectors are have superarrows (vector $\vec{a}$) while scalars are $italic$: scalar $a$.}; the consequent minus sign can be interpreted to be an operation (remove $\vec{a}t$ from $\vec{v_i}$) or as a comparison between relative directions (of $\vec{v_i}$ and $\vec{a}$) or simply as a negative acceleration (as in the speed model).  Students may be unaware of the multiple meanings, or may flexibly switch meaning implicitly \cite{hayes2010role}, confusing themselves in the process.  Students sometimes treat the acceleration symbol $\vec{a}$ as a constant which is always positive and used the ``outer minus'' to obtain its negative value. In other cases, the student treats the symbol as a variable which could contain the ``inner minus'' and be negative itself, leading to inconsistent solutions \cite{hayes2010role}. 

Recent studies \cite{brahmia2016exploring,Brahmia2018} make efforts to investigate student understanding of negativity and positivity in introductory physics contexts in a more systematic manner. Starting from the claim that the negative sign could account for multiple meanings, the authors used the Vlassis' categorization scheme to ask students to explicitly explain what physical meaning is embodied in individual negative and positive quantities in different mechanics and electromagnetism contexts. The results show that even though students' flexibility in interpreting the meanings of signs is highly context-dependent, they generally struggle with the symmetrical meaning more than unary and binary functions. Moreover, combining more than one nature of negativity in a single calculation appears to present significant challenges to introductory students. 

Altogether, these disparate accounts of signs in mathematics and physics suggest that students struggle with flexibly choosing and applying specific meanings of signs, from children to intermediate-level undergraduate physics students.  This is not surprising: there are multiple meanings of the negative sign which may vary among problems and within a given problem or expression.  

\section{\label{sec:TheoreticalFramework}Theoretical Framework}

We turn conceptual blending theory to help us make sense of the different meanings of signs in physics.  

\subsection{\label{subsec:ConceptualTheory}Conceptual blending theory}

\begin{figure}[hptb]
  \includegraphics[width=0.8\linewidth]{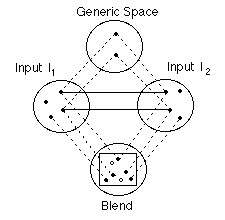}
  \caption{Mental network model \cite{fauconnier2008way}}
  \label{fig:mentalnetwork}
\end{figure}

Conceptual blending theory was developed by \citet{fauconnier2008way} to account for how people create meaning. The theory posits a mental network model that processes and forms new meaning (Figure \ref{fig:mentalnetwork}). The mental network model is composed of at least two input spaces containing information from discrete domains, a generic space containing common information and structures, and a blended space where new meaning emerges. The input elements in one space connect to their counterparts in the other spaces via vital relations, such as time, space, change, cause-effect, identity, etc. When projected into the blended space, these vital relations could be compressed into the same types, or more often, different types of relations in the blended space.

The process of generating new meaning in the blended space involves three operations: composition, completion, and elaboration. Composition sets up the input spaces and the relation between them. Completion is where the conceptual structure and knowledge from long term memory are recruited to complete the composed structure. Elaboration is where the blend is developed through an imaginative simulation according to the new principles and structure in the blend, and possibly creates other new principles and structures in the blend itself consequently. In other words, new meaning is created via a process of setting up the mental spaces, matching across the spaces, locating shared structures, projecting input elements into the blend, projecting backward to the inputs, recruiting new structures to the inputs or the blend, and running various operations in the blend itself. This is usually a fast process, happening at a subconscious level.

In the mental network model, the projection of elements from input spaces into the blended space is selective, which means not all elements and structures in input spaces will be projected into the blended space. 
Selective projection allows multiple blends with different imaginative networks to be constructed from the same two input spaces. This is largely overlooked in the literature. Indeed, researchers have built a substantial number of rigorous blends, showing the construction of meaning in the context of everyday life where the given input spaces mostly produce only one blended space with specific and predictable meaning. In this study, we emphasize the possibility that the same input spaces can produce different meanings due to different ways of projecting input elements into the blended space. Therefore, even though constructing blends might be fast, easy, and subconscious, selecting an effective blend or emergent meaning for a specific context could be more difficult.

\subsection{\label{subsec:ConceptualTheory2}Conceptual blending theory in science education research}
The theory of conceptual blending has been extensively applied and studied in many fields and plays a central role in discovering and developing new mathematical ideas. In physics education research, studies have focused on choosing appropriate inputs and building blended spaces to explore how students understand specific concepts such as energy \cite{dreyfus2015applying} and waves \cite{wittmann2010using}. For instance, the idea of a wave propagating as an object comes from blending a wave input with a ball input whereas the idea of a wave propagating as an event comes from blending a wave input with a domino input. Other works investigate how students blend among different physical representations, such as sound waves, string waves, and electromagnetic waves \cite{podolefsky2006use}; how students blend hand gestures when reasoning about mathematical ratios \cite{edwards2009gestures}; and how students use arrows to mean vector quantities in the context of electric fields \cite{gire2014arrows}.  

Research uses blending theory to investigate the ways students blend mathematics and physics together and make sense of the physical world as well. Studies \cite{bing2007cognitive, hu2013using} have attributed students' struggles in using mathematics not to their lack of prerequisite knowledge and skills but to their inappropriate mapping or blending between mathematics and the physical world.  For instance, a student who frames a physics problem as only a mathematical one might wind up on the wrong track for an extended period of time, even when that student excels at mathematical reasoning \cite{redish2014oersted}. Therefore, the effective use of mathematics in making sense of the physical world involves blending reciprocally between mathematics and physics contexts rather than just applying mathematics to physics. 


These diverse studies of the conceptual blending theory in physics education research have investigated how students choose input spaces, project ideas forward to a blended space, and operate within a blended space.  They tend to focus on how different input spaces can create different blends \cite{dreyfus2015applying, podolefsky2006use, gire2014arrows, wittmann2010using}.  The process of making meaning with blends can be effortful and fraught\cite{redish2014oersted, bing2007cognitive, hu2013using, gire2014arrows}, and careful curriculum development can guide students to choose appropriate inputs and elaborate appropriate blends\cite{podolefsky2006use, redish2014oersted}.

In our work, we bring forward ideas from both literatures.  From cognitive linguistics, we take up the idea that the same input spaces can produce multiple blends via selective projection, but that the process of blending can be very fast and unconscious.  From physics education research, we take up the idea that making meaning with blends can be effortful, and that coordinating ideas from multiple representations (particularly algebraic) can be difficult.  We attribute students' struggles to solve a problem to the difficulty in choosing appropriate blends from the same input spaces, not to difficulty choosing input spaces or difficulty running the blend. Our argument is driven by theory and supported with case-study observational data; recommendations for how to teach this material are outside the scope of this paper.

\section{\label{sec:Problem}Problem of Interest and General Solution}

\begin{figure}[hptb]
  \includegraphics[width=\linewidth]{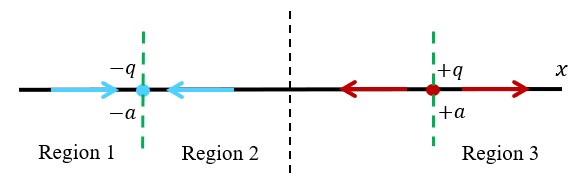}
  \caption{Electric field contributions caused by the charges $-q$ (light blue arrows on the left) and $+q$ (dark red arrows on the right) along the $x$ axis.}
  \label{fig:efields}
\end{figure}

We ground our argument in one problem from introductory electrostatics.  
Suppose we have two charges at $-a$ and $+a$ on the x-axis as in Figure \ref{fig:efields}. What does the electric field look like along the x-axis? 

The electric field contributions from both charges change in direction and vary in magnitude as you move along the axis. These contributions are commonly represented with arrows such as those shown in Figure \ref{fig:efields}. 

Coulomb's law gives the electric field $\vec{E}$ caused by a point charge $\pm q$ as $\vec{E}=k\frac{\pm q}{R^2} \hat{R}$, where $k$ is a (positive) constant. In this equation, $\vec{R}$ is the vector distance pointing from the point charge to a field point along the axis, where $R_{+q} = x-a$ and $R_{-q} = x+a$.   Electric field contributions from different charges superpose to yield the electric field at any point. 
%
%
%
 
You could divide the axis into three regions -- left of both charges, right of both charges, and between the two charges -- and consider each region separately.  In this approach, one can flexibly take the magnitudes of the field contributions $E = |k\frac{ (\pm q)}{R^2}|$, project the electric field direction on the $x$ axis, and obtain the following solutions for each region as:

\begin{align}
\text{General:}\nonumber\\
\vec{E} &= \vec{E}_{+q} + \vec{E}_{-q}\label{eqn:general}\\
\text{Region 1:} \nonumber\\
 \vec{E}_1 
 		&= k\frac{ q}{(x-a)^2} (-\hat{x}) + k\frac{ q}{(x+a)^2} (+\hat{x}) \\
\text{Region 2:} \nonumber\\
 \vec{E}_2 
 		&= k\frac{ q}{(x-a)^2} (-\hat{x}) + k\frac{ q}{(x+a)^2} (-\hat{x}) \\
 \text{Region 3:}\nonumber \\
 \vec{E}_3 
 		&= k\frac{ q}{(x-a)^2} (+\hat{x}) + k\frac{ q}{(x+a)^2} (-\hat{x}) 
 \end{align}


Other methods are possible; however, solving the problem correctly requires simultaneous consideration and consistency in treating directionality, algebraic values of location, and distance in the denominators. Using conceptual blending theory, we investigate the multiple meanings that might be associated with the signs in this problem.

\section{\label{sec:blends}Blends of interest}
In the context of this electric field problem, we claim that the signs carry two domains of meaning. First, the electric field is a vector quantity, so the signs bear the meaning of directionality. Second, the electric field expression involves the calculation of distance between the point charge and the field point, which depends on how one accounts for the locations of these points relative to the origin and to each other. 

Thus, we have two different input spaces to blend with algebraic \textit{sign}: \textit{directionality} and \textit{location}.  Depending on which input elements are used, the same two input spaces may produce different blends, which could then lead to different conclusions about the physics involved. 


\subsection{Blending between directionality and sign}
\begin{figure*}[bt]
  \includegraphics[width=\linewidth]{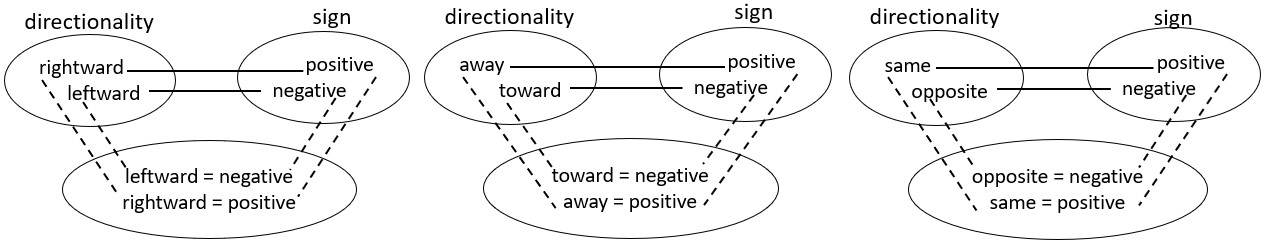}
  \caption{From left to right: space-fixed blend ($\square$), body-fixed blend ($\triangle$), and comparative blend ($\Circle$).}
  \label{fig:spaceblends}
\end{figure*}

We propose three different blends between directionality and algebraic sign\cite{huynh2018blending}. In each blend, there are two input spaces - \textit{directionality} and \textit{sign}. The \textit{directionality} input can contain elements expressing direction in one dimension such as rightward, leftward, away, toward, same, opposite, up, down, etc. The \textit{sign} input contains the elements of negative and positive. The blending between the first three pairs of these \textit{directionality} elements and these \textit{sign} elements makes space-fixed, body-fixed, and comparative blends, respectively (symbolically referred as shown in Figure \ref{fig:spaceblends}). The elements of up and down are not applicable in this horizontal system, and hence their blend is not shown. 

A space-fixed blend ($\square$) occurs when rightward and leftward from \textit{directionality} are projected into the blended space. This blend is common and appropriate in a space-fixed problem, such as when there is a coordinate axis in a one-dimensional problem. This axis could be positive leftward or rightward. From \textit{directionality}, rightward and leftward map to positive and negative (respectively) from \textit{sign}, projecting forward the convention that leftward is negative and rightward is positive. Running the blend yields a unit vector (e.g. $\hat{x}$) which is positive when it points to the right as a convention for one-dimensional, space-fixed coordinate systems in physics.

Alternately, one could select away and towards from \textit{directionality} to map to positive and negative in \textit{sign} (respectively). This is common in body-fixed coordinate systems: moving away from me is positive velocity and moving toward me is negative velocity in person-centric motion \cite{Tabachnick2018}; radial vectors are positive away from the source. Consider the equation that expresses Coulomb's law for the electric field caused by a point charge,  $\vec{E} = k\frac{ \pm q
}{R^2} \hat{R}$. Because the distance vector $\hat{R}$ always points away from the point charge, the positive sign accounts for the pointing-away electric field in the blended \textit{directionality-sign} space, and negative sign accounts for the pointing-toward electric field accordingly. This meaning is established in electromagnetism and requires a connection to the meaning of the distance vector $\hat{R}$. We see students implicitly and unconsciously refer to the body-fixed blend ($\triangle$) when making charges carry signs, rather than discussing the inward and outward directions of electric fields. 

Another emergent meaning of sign comes from the blend of \textit{sign} with the relative direction of two vectors where the characteristics of \textit{directionality} are now sameness and oppositeness. This comparative blend ($\Circle$) might originate formally from the mathematical property of the inner product of two vectors; the inner product of two opposite vectors is negative. Apart from comparing the signs of pairs of vectors, such as $\hat{x}$, $\hat{E}$, and $\hat{R}$, the meaning that emerges from this comparative blend also sometimes shows up when students consider the interference of two fields. For instance, $\vec{E}_1$ and $\vec{E}_2$ are destructive if their directions are opposite. Thus, one might insert an outer negative sign accordingly to account for that destructiveness in the expression for the total field strength. Note that constructiveness and destructiveness come from relative direction, which is represented by the inner sign of the vector quantity. If students treat direction and magnitude separately, they might bring both inner or outer signs into the direction comparison and double account for the meaning of relative direction.

A quick typographical note is in order.  We have introduced the symbols $\triangle$ (body-fixed blend), $\Circle$ (comparative blend), and $\square$ (space-fixed blend) for the three blends between \textit{directionality} and \textit{sign}.  We're about to introduce the symbols $\blacksquare$ (signed blend) and $\blacktriangle$ (unsigned blend) for two blends between \textit{location} and \textit{sign}.  We chose the fill of these symbols to connect to \textit{directionality} (open) and \textit{location} (filled); the actual shapes are for typographical ease.  Compressing the names of the blends into the symbols allows us to compactly describe a series of blends in problem solving, which is important to the work of this paper.  

\subsection{Blending between location and sign}

\begin{figure*}[bt]
  \includegraphics[width=0.75\linewidth]{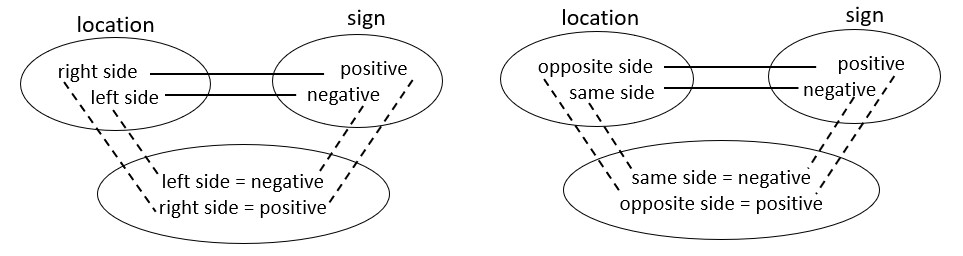}
  \caption{From left to right: signed blend ($\blacksquare$) and unsigned blend ($\blacktriangle$).}
  \label{fig:signblends}
\end{figure*}

\begin{figure}[hpb]
  \includegraphics[width=\linewidth]{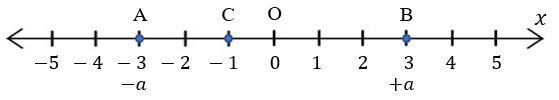}
  \caption{Axis and number line}
  \label{fig:axis}
\end{figure}

In addition to the three blends between \textit{directionality} and \textit{sign}, we construct two blends between \textit{location} and \textit{sign}, to fully cover the problem at hand. The \textit{location} input can contain elements expressing location in one dimension such as right of origin, left of origin, opposite side, and same side. Blending between pairs of these location elements and the positive and negative sign elements makes signed and unsigned blends, respectively (symbolically referred as shown in Figure \ref{fig:signblends}). 

In the signed blend, the location of a point to the left or right of the origin maps to negative and positive signs respectively. This blend yields the signed coordinate of a point on an axis consistent with the conventional choice of rightward-positive axis. For example, in Figure \ref{fig:axis}, one can read out the coordinates of points $A$ and $B$ accordingly as $x_A=-a$ and $x_B=+a$. In this example, the letter $a$ is a positive, constant, ``unsigned'' number.  However, the symbols $x_A$ and $x_B$ could carry signs inside them and be positive or negative: they are ``signed'' numbers. Unsigned and signed numbers are also termed ``positive constants'' and ``variables'' (respectively) \cite{hayes2010role}, though we note that other definitions of variable don't exactly match our sense of signed number: one can take a derivative with respect to a variable, but not with respect to a signed number. When taking up the signed blend, the distance between two points is drawn from the formal mathematical operation of \textit{difference}. For example, the distances $AC$ and $BC$ as shown in Figure \ref{fig:axis} are $AC=|x_C-x_A|=|-1+3|=2$ and $BC=|x_C-x_B |=|-1-3|=4$. This blend showed up in our data when the students checked that the difference in location between two points is zero when those two points are on top of each other.

On the other hand, one could treat coordinates as unsigned numbers and view them as the length of some distance from the origin.  Under the unsigned blend (Figure \ref{fig:signblends}), the total distance between two points is subtracted when the two points are on the same side of the origin and added when the two points are on opposite sides of the origin. For instance, the distances $AC$ and $BC$ in Figure \ref{fig:axis} are found as $AC=|-3|-|-1|=2$ and $BC=|3|+|-1|=4$. 

The signed and unsigned blends independently give the same answers for the distance between two points. However, the blends require different thinking about the nature of coordinates, and therefore different operations to find the distances.  The signed blend is especially useful and necessary when the coordinates of the points of interest are undefined whereas the unsigned blend is more useful when all coordinates are well-defined. 

In the particular case of positive signed numbers, distinctions between the corresponding blends for variables and constants is unnecessary. Taking the difference of two signed numbers located on the same side of the origin (signed blend) is formally the same as and interchangeable with subtracting those two lengths (unsigned blend). Otherwise, one must be consistent in using one or the other approach, or else errors will  occur.

Even though multiple blends from the same input spaces is permitted in conceptual blending theory via the process of selective projection, this aspect of the theory has been largely ignored in the science education literature.  It could be that choosing among these different blends from \textit{sign} and \textit{directionality} and from \textit{sign} and \textit{location} contributes to what makes working with positive and negative sign hard for students. Although running these blends is fast and subconscious, choosing an appropriate blend or flexibly shifting the meaning of the sign is complicated, and hence hard.

\subsection{Exemplar solution with blending}
We illustrate multiple ways to solve the problem with the blends we build. We start with the body-fixed blend ($\triangle$), which is embedded in Coulomb's law (Table \ref{tab:exemplarsoln}, line 1). 
For each charge, we get:
\begin{align}
\vec{E}_{+q} &= k\frac{(+q)}{R_+^2 } \hat{R}_+ \\
\vec{E}_{-q} &= k\frac{(-q)}{R_-^2 } \hat{R}_-
\end{align}

In this solution, we will treat the signs of $\vec{E}_{+q}$ in each region using blends between \textit{directionality} and \textit{sign} in two ways: first, with the space-fixed blend ($\Square$) and then with the comparative blend ($\Circle$), arguing that similar blends can be made for $\vec{E}_{-q}$.  Then, we will blend \textit{location} and \textit{sign} to consider the denominators in terms of $x$ and $\pm a$.  

\begin{table*}[tb]
\begin{tabular}{|c|p{2cm}|c|l|p{6cm}|}
\hline
Line & Region  & Blend      & Reason                             & Math                              \\ \hline
1	&	General	& $\triangle$ &Coulomb's Law &	$\vec{E}_{+q} = k\frac{(+q)}{R_+^2 } \hat{R}_+$ ; $\vec{E}_{-q} = k\frac{(-q)}{R_-^2 } \hat{R}_-$ \\ 
\hline
\hline
2	&	1\&2 	&	$\Square$ &$\vec{E}_{+q}$ points left & $\vec{E}_{+q}$ is negative\\
3	&  3 	&	$\Square$ &$\vec{E}_{+q}$ points right & $\vec{E}_{+q}$ is positive\\
\hline
4	&	1\&2 	&	$\Circle$ &$\hat{E}_{+q}$ and $\hat{x}$ oppose & $\hat{E}_{+q} = -\hat{x}$\\
5 	& 	3	&	$\Circle$& $\hat{E}_{+q}$ and $\hat{x}$ parallel & $\hat{E}_{+q}=+\hat{x}$ \\
 \hline
 6	&	1 	&	$\Square$ &$\vec{E}_{-q}$ points left & $\vec{E}_{-q}$ is positive\\
7	&  2\&3 	&	$\Square$ &$\vec{E}_{-q}$ points right & $\vec{E}_{-q}$ is negative\\
\hline
8	&	1 	&	$\Circle$ &$\hat{E}_{-q}$ and $\hat{x}$ oppose & $\hat{E}_{-q} = \hat{x}$\\
9 	& 	2\&3	&	$\Circle$& $\hat{E}_{-q}$ and $\hat{x}$ parallel & $\hat{E}_{-q}=-\hat{x}$ \\
 \hline
 \hline
 10 & 1\newline 2\newline 3 & &Superposition&{$ \vec{E}=(-E_{+q} + E_{-q})(+\hat{x}) \newline
							\vec{E} =(E_{+q} + E_{-q})(-\hat{x}) \newline  
 							\vec{E} =(E_{+q} - E_{-q})(+\hat{x})$} \\
\hline
\hline
11 & General & $\blacksquare$ &$x$, $x_{-q}$, and $x_{+q}$ are signed numbers & $R_+ = x - x_{+q}~ = x-(+a) = x-a$ \newline
																	$R_- = x - x_{-q}~ = x-(-a) = x+a$ \\
																	\hline
12 & left of origin & $\blacktriangle$ & $x$, $x_{-q}$, and $x_{+q}$ are unsigned numbers &$R_+ = x+ a$\newline $R_- = x-a$ \\
13 & right of origin & $\blacktriangle$ & $x$, $x_{-q}$, and $x_{+q}$ are unsigned numbers  & $R_+ = x-a$\newline $R_- = x+ a$ \\

\hline
\end{tabular}
\caption{Exemplar solution. Space-fixed blend = $\square$. Body-fixed blend =$\triangle$.  Comparative blend =$\Circle$. Signed blend = $\blacksquare$. Unsigned blend =$\blacktriangle$.} \label{tab:exemplarsoln}
\end{table*}


First we determine the direction of $\vec{E}_{+q}$ and $\vec{E}_{-q}$ using the space-fixed blend ($\Square$). For instance, in regions 1 and 2 in Fig. 2, $\vec{E}_{+q}$ points to the left and hence is negative (Table \ref{tab:exemplarsoln}, line 2). In region 3, however, $\vec{E}_{+q}$ points to the right and is therefore positive in that region (Table \ref{tab:exemplarsoln}, line 3). 

Alternatively, we can explicitly compare ($\Circle$) $\hat{E}_{+q}$ with $\hat{x}$. Then, $\hat{E}_{+q}$ and $\hat{x}$ are in opposite directions in regions 1 and 2, and therefore $\hat{E}_{+q} = -\hat{x}$ in these regions (Table \ref{tab:exemplarsoln}, line 4). $\hat{E}_{+q}$ is parallel to $\hat{x}$ in region 3, however, and thus $\hat{E}_{+q}=+\hat{x}$ in this region (Table \ref{tab:exemplarsoln}, line 5). Both approaches correctly lead to the same expression for $\vec{E}_{+q}$ in each region
, but the blends are different.  The difference between them is subtle but can be distinguished in most cases through observations of students' word choices, diagrams, and gestures.  Similar reasoning yields the signs for $\vec{E}_{-q}$ in all three regions (Table \ref{tab:exemplarsoln}, lines 6-9).

To find the total field, the student might use the superposition theorem, where the signs of the component fields are consistent with their relative directions: opposite $\rightarrow$ negative, same $\rightarrow$ positive ($\Circle$) (Table \ref{tab:exemplarsoln}, line 10).


The complete solution requires the determination of $R_+$ and $R_-$ in terms of $x$ and $\pm a$.  There are two possible blends here with \textit{location} and \textit{sign}.  Using the signed blend ($\blacksquare$),  $x$, $x_{-q}=-a$, and $x_{+q}=+a$ are signed numbers. The differences between them are given in Table \ref{tab:exemplarsoln}, line 11. Alternately, you could use the unsigned blend ($\blacktriangle$) to yield  $x$ and $x_{-q}=x_{+q}=a$ as unsigned numbers and find the distance between them with the sense of length (Table \ref{tab:exemplarsoln}, line 12\&13). While the signed blend ($\blacksquare$) approach gives the same value for $R_+$ and $R_-$ along the $x$ axis, the unsigned approach ($\blacktriangle$) presents additional difficulties at the origin. Although the two approaches will reveal the same physical properties, using the signed blend ($\blacksquare$) is more proper.

Altogether, we have articulated two possible blending paths for the directionality of $\vec{E}_-$ and $\vec{E}_+$ (using $\Circle$ and $\Square$) and two possible blending paths for the denominators (using $\blacktriangle$ and $\blacksquare$).  In combination, this is four solution paths to this problem: 
\begin{align}
\triangle \rightarrow& \Circle\rightarrow \Circle \rightarrow \blacktriangle \rightarrow \blacktriangle \label{eqn:exemplar1}\\
\triangle \rightarrow& \Circle \rightarrow\Circle \rightarrow \blacksquare \\
\triangle \rightarrow& \Square \rightarrow \Square\rightarrow \blacktriangle \rightarrow \blacktriangle \\
\triangle \rightarrow&  \Square \rightarrow \Square \rightarrow\blacksquare \label{eqn:exemplar4}
\end{align}


\section{Instructional Context}
The data is drawn from an upper-division Electromagnetism I course, taught in the Fall 2013. The course enrollment was around 20 students and was taught by a white female instructor with experience. The course covered the first seven chapters of \textit{Introduction to Electrodynamics} ($4^{th}$ edition) by David J. Griffiths. Throughout the semester, the class met four times per week for fifty minutes each, during which the students solved tutorials or problems in groups of three or four and interacted with the instructor intermittently. The class instruction focused on how physics and mathematics work together and on the process of exploring physical systems, and not on memorizing procedures and formulas.  The grading criteria for this course placed emphasis on the process of physical sense-making and mathematical reasoning and checking. Additional details on how the course was structured and students' activities during class are available in other papers\cite{Chari2019InsFraming, Nguyen2016StuFraming}; this paper does not dwell on the instructional details of the course. The students in this course had relatively strong mathematical backgrounds and mathematical sense-making skills compared to introductory students. 

Two oral exams were mandatory for all students. One covered problems with electric fields, and the other covered problems with magnetic fields. Each oral exam lasted approximately 30 to 40 minutes, and each student was encouraged to think-out-loud as they solved the problem on the whiteboard; almost all oral exams were video-recorded. The class was videotaped throughout the semester and hence, most of the students felt comfortable with working in front of a camera. The data is drawn from the first oral exam that the students took individually during the fourth week of the course. This was after the class had covered major topics in electrostatics such as Coulomb's law, Gauss's law, conductors, and the method of separation of variables. Because we are interested in students' conceptual understanding of mathematics meaning in physics problem solving, student oral exams are an appropriate source for data selection \cite{sayre2014oral}.

Four male students (three white, one Asian) were given this problem as part of their individual oral exam. The problem was more thought-provoking for these students than one might initially assume, due to their struggles with making sense of the signs coming from multiple sources. Indeed, none of them achieved success in the first few tries.

\section{Methodology}
When investigating the data, we looked for evidence that students employ multiple blends and switch among those blends to handle the given problem. Since the input elements of blends are fairly distinct in most cases, we could easily recognize and differentiate students' choice of blends via their verbal reasoning, gesture, diagram, and mathematical representations. Indeed, as the problem involves multiple positive and negative signs with different meaning, we observed that our students often put negative signs next to or in the brackets with the quantities that they think the signs belong to. Yet, those students with relatively strong algebraic backgrounds sometimes assisted their reasoning with formal mathematics and skipped some intermediate steps on the board. Therefore, it is important to attend to students' coordination among multiple representations. 

For instance, accounting for the charge signs in the equation indicates the body-fixed blend. The words \textit{pushing that way}, or \textit{$\vec{E}_{+q}$ points in $-\hat{x}$} along with a constant pointing gesture indicate the direction of left or right, and thus the space-fixed blend. The words \textit{coming destructively} or \textit{$\vec{E}_{+q}$ and $\vec{E}_{-q}$ have opposite signs} along with gestures or diagrams that represent two opposing objects reveal the comparative blend. Vaguely stated reasoning, such as \textit{$\vec{E}_{+q}$ is negative}, is insufficient when attempting to draw conclusions about students' input spaces, and hence their blends. Such reasoning might or might not be clarified with other representations' assistance.

Verbal reasoning also provided us with indications of students' metacognitive awareness. Metacognition refers to thoughts about thoughts or reflections upon one's actions \cite{tsai1998analysis}. The study of metacognition has developed across disciplines and has shown a positive correlation between students' performance and their metacognitive ability \cite{Vestal2017}. Recent work \cite{sayre2015brief} investigated how metacognitive talk could reveal students' thinking-like-a-physicist behavior. Three types of metacognitive talk were identified in our analysis: understanding, confusion, and spotting inconsistency. Even though we are not interested in thinking-like-a-physicist behavior here, paying attention to students' metacognitive awareness hints at their choice of blend actions and reasoning: picking blends, checking blends, and changing blends. Therefore, we prefer the pieces of data where students are metacognitively good at expressing their thoughts as they assign meaning to signs.

We transcribed the data and wrote narratives with detailed descriptions of students' coordination between verbal reasoning and other presentations, such as mathematics and gesture. Along with the video, we broke students' reasoning into mathematical steps. This was sensible because students usually completed their choices on the roles of signs temporarily as they worked through those steps. We mapped students' ideas in each step to the input elements of blends, and then converted the meaning they associated with signs into corresponding blends. In some steps, we observed the students employ more than one of the blends, which was also appropriate since multiple signs can simultaneously appear in a single mathematical expression. Significantly, we will consider instances when students recognize conflicts between ideas of positive and negative signs with different meanings, deploy some of those meanings, or affiliate one meaning with another meaning, to be supporting our claims.  Two researchers reviewed all of our blending results; a committee of three additional researchers reviewed a subset and came to consensus on them.

In the next sections, we will analyze two case-studies within our data. With overall differences in approach, these two cases help us illustrate the existence of the multiple blends that students can construct and employ when attempting to solve the problem. 

\section{Oliver Case Study}

Among four students solving the same problem of interest, we choose to first look at Oliver's reasoning with signs. Although Oliver did not struggle much with signed and unsigned numbers, his reasoning with signs and directionality was typical among our students. Oliver usually tried to express his mathematics in a manner consistent with the way that he thought about physical systems, i.e., he put the sign in the bracket with the variables to which he thought that sign belonged. He was also good at thinking aloud, which helped us to understand his thoughts on signs and the blends he was using. Part of this case study is given in detail in our previous work \cite{huynh2018blending}. To compare with the other case study in this paper, we give full details on this analysis again here.

\begin{table*}[tb]
\begin{tabular}{|c|p{2cm}|c|l|p{6cm}|}
\hline
Move & Region  & Blend      & Reason                             & Math                 \\
\hline
1	&	General & $\Circle$ & $\hat{R}$ and $\hat{x}$ parallel  & $\hat{R} = \hat{x}$ \\
    &   & $\blacksquare$    & Definition of $\vec{R}$ & $R_{-q} = x+a$, $R_{+q} = x-a$\\
    &	& $\triangle$ & Coulomb's Law  &	 $\vec{E} =\vec{E}_{-q}+\vec{E}_{+q}=k\frac{-q}{(x+a)^2} \hat{x}+k\frac{q}{(x-a)^2} \hat{x}$  \\
\hline
2   &   General & $\Square$ & $\vec{E}_{+q}$ and $\vec{E}_{-q}$ could point left & $\vec{E}_{+q}$ and $\vec{E}_{-q}$ could be negative    \\
\hline
\hline
3	&	1 	&	$\Circle$ & $\vec{E}_{+q}$ and $\vec{E}_{-q}$ destructive & $\vec{E}_1=\vec{E}_{-q}-\vec{E}_{+q}=k\frac{-q}{(x+a)^2} \hat{x}-k\frac{q}{(x-a)^2} \hat{x}$    \\
\hline
4	&  1 	&	$\Square$ & $\vec{E}_{1}$ points right & $\hat{E}_{1} = +\hat{x}$\\
\hline
5	&   1 	&	$\Circle$ &$\hat{E}_{+q}$ and $\hat{E}_{-q}$ oppose & $\vec{E}_1=k{[}\frac{-q}{(x+a)^2}+\frac{q}{(x-a)^2}{]} \hat{x}$\\
\hline
6 	& 	1	&	$\Square$   & $\hat{E}_{+q}$ points left, $\hat{E}_{-q}$ points right & $\vec{E}_1=k{[}\frac{q}{(x+a)^2}-\frac{q}{(x-a)^2}{]}\hat{x}$ \\
 \hline
 \hline
7   & 2  & $\Square$ & $\hat{E}_{+q}$ and $\hat{E}_{-q}$ point left & $\hat{E}_{+q} = \hat{E}_{-q} = -\hat{x}$ \\
    &   & $\triangle$ & Coulomb's Law & $\vec{E}_2=\vec{E}_{-q}+\vec{E}_{+q}=k\frac{-q}{(x+a)^2}(-\hat{x})+k\frac{q}{(x-a)^2}(-\hat{x})$ \\
\hline
8	&	2 	&	$\Circle$ &$\hat{E}_{+q}$ and $\hat{E}_{-q}$ parallel & $\hat{E}_{-q} = \hat{E}_{+q}$    \\
    & 		&	$\square$ & $\hat{E}_{+q}$ and $\hat{E}_{-q}$ point left & $\vec{E}_2=\vec{E}_{-q}+\vec{E}_{+q}=k{[}\frac{q}{(x+a)^2}+\frac{q}{(x-a)^2}{]}(-\hat{x})$ \\
 \hline
 \hline
9 & 3 & $\Circle$ & $\vec{E}_3$ and $\vec{E}_1$ oppose &  \\
    &   & $\square$ & $\hat{E}_{+q}$ points right, $\hat{E}_{-q}$ points left & $\vec{E}_3=\vec{E}_{-q}+\vec{E}_{+q}=k{[}\frac{-q}{(x+a)^2}+\frac{q}{(x-a)^2}{]}\hat{x}$ \\

\hline
\end{tabular}
\caption{Oliver derives solution. Space-fixed blend = $\square$. Body-fixed blend =$\triangle$.  Comparative blend =$\Circle$.} \label{tab:oliver}
\end{table*}

Oliver started by recording the superposition formula of the general total electric field $\vec{E}$ = $\vec{E}_{-q}$ + $\vec{E}_{+q}$, and then defined each contribution using Coulomb's law (Table \ref{tab:oliver}, move 1), where the body-fixed blend ($\triangle$) is embodied in each charges' sign. Oliver spent a few minutes to recall the definition of $\vec{R} = \vec{r} - \vec{r'}$ to determine distance $R_{+q}$ and $R_{-q}$, which suggests that the signed blend was used. Throughout his work, the denominators $R_{-q}=x+a$ and $R_{+q}=x-a$ were persistently ascribed to $\vec{E}_{-q}$ and $\vec{E}_{+q}$, respectively (Table \ref{tab:oliver}, move 1).

Despite of the different values of $R_{+q}$ and $R_{-q}$, Oliver treated their directions as separate from their magnitudes and equal: $\hat{R}_{+q}$ = $\hat{R}_{-q}$ = $\hat{R}$. Indeed, Oliver first thought that $\hat{R}$ depended on where he picked the field point but then he decided to just change into $\hat{x}$ since ``we [consider] the whole x axis''. For illustration, a positive sign commensurate with the \textit{sameness} ($\Circle$) of the $\hat{R}$ and $\hat{x}$ directions was added for his replacement reasoning (Table \ref{tab:oliver}, move 1). This incorrect comparison between $\hat{R}$ and $\hat{x}$ lead Oliver to an expression that conflicted with other blends that he would later use.``But it can be \textit{negative} [pointing \textit{left}]'' - Comparing with the space-fixed blend ($\Square$) helped Oliver to notice that his math did not satisfy the variation in the electric field direction along the $x$ axis (Table \ref{tab:oliver}, move 2). He then defined the field vectors on the diagram and decided to divide the given region into three smaller ones (Fig. \ref{fig:efields}).

In region 1, Oliver inserted a negative sign between the two terms because ``they are going to be \textit{destructive} (\Circle)\dots Out here, [$\vec{E}_{-q}$] has greater effect than the contribution from [$+q$ charge][\dots ] It is going to be [$\vec{E}_{-q}$] \textit{minus} [$\vec{E}_{+q}$]'' (Table \ref{tab:oliver}, move 3). While recording the result with two negative terms, he recognized ``it doesn't seem right'' due to the conflicts with his earlier reasoning of the total field ``[$\vec{E}_1$] is going to \textit{face to the right} [pointing \textit{right}] ($\Square$), [\dots ] \textit{plus} $\hat{x}$'' (Table \ref{tab:oliver}, move 4). Oliver sensed that fields should have \textit{opposite signs} when one field tends to reduce the other but in his math showed that ``[\dots] they are both negative, [\dots] it looks like they kind of add together''. Therefore, he decided to deploy the destructiveness meaning of sign and changed the second negative term back to positive to satisfy the relative opposing directions ($\Circle$) of $\vec{E}_{-q}$ and $\vec{E}_{+q}$ (Table \ref{tab:oliver}, move 5). Oliver did this without regard to the direction of each electric field component in relation to $\hat{x}$.

From his mathematical expression, the instructor pointed out that $\vec{E}_{+q}$ was now pointing in the $+\hat{x}$ direction. He became frustrated manipulating the multiple signs appearing with the different meanings of directionality. Oliver was certain that ``[$\vec{E}_{+q}$] \textit{points in} $-\hat{x}$ direction out here, and then [$\vec{E}_{-q}$] \textit{points in the positive} right here ($\Square$), because it is pointing in'' as shown in the diagram. He changed the signs of the terms accordingly (Table \ref{tab:oliver}, move 6) and explained his thoughts on the final signs' meaning: ``Ok, I know what I did wrong\dots  Because\dots  see the charges, I should have just figured it out or thought about which direction it is. This is exactly what is changing the signs, not necessarily the \textit{sign of the charges}.'' Clearly, Oliver had not figured out that the root of all conflict was in the comparison of $\hat{R}$ and $\hat{x}$, but he was beginning to understand that the body-fixed blend ($\triangle)$ was not as useful to him as the space-fixed ($\Square$) and comparative ($\Circle$) blends .

Although Oliver had assigned appropriate directionality meaning to the signs in the electric field expression in region 1, he failed to repeat this reasoning in region 2. Oliver first reasoned that ``[the fields] are going to be constructive [\dots] and by constructive I mean they are both going in the same direction [pointing \textit{left}] $\Square$'' . He also blended the effect of the charge signs ($\triangle$) into his math. For example, ``[for the charge $-q$] it is \textit{minus} charge ($\triangle$) but it can have \textit{negative $\hat{x}$ direction} [\dots ] since it's \textit{backwards} [pointing \textit{left}] ($\Square$)'' (Table \ref{tab:oliver}, move 7). ``That doesn't make sense'' - Oliver got stuck when he contrasted: ``[They should be in] the \textit{same} direction ($\Circle$) [\dots ] \textit{to the left} ($\Square$), so $-\hat{x}$ [\dots ] Yea, I am confused''. Oliver again struggled with affiliating multiple meanings to signs. As the instructor reminded him of his correct reasoning about signs in region 1, Oliver reminded himself to ``not worry about the sign [of charge], just worry about the field [direction]''. Oliver was now ready to adjust his expression by deploying the body-fixed blend. He wrote down expression from his earlier reasoning, which involved the comparative ($\Circle$) and spaced-fixed ($\Square$) blends (Table \ref{tab:oliver}, move 8).

In region 3, Oliver quickly recognized that ``[$\vec{E}_3$] is going to be much like the [field in] region 1, except exactly the \textit{opposite} ($\Circle$)''. As he recorded the solution (Table \ref{tab:oliver}, move 9), Oliver clearly stated that ``[$\vec{E}_{-q}$] is going to be \textit{negative $\hat{x}$} [pointing \textit{left}], and [\dots $\vec{E}_{+q}$] is going to be positive $\hat{x}$, because\dots  so for this charge [$-q$], the field is pointing \textit{inwards} [pointing \textit{left}] and for this charge [$+q$], it's pointing \textit{out} [pointing \textit{right}] ($\Square$).'' Despite Oliver's use of the words \textit{inwards} and \textit{outwards}, we argue that Oliver might not have been referring to the body-fixed blend. First, he arrived at the solution initially by gesturing and by mathematically expressing the left - right direction; the explanation was to reaffirm the directions on the diagram. Second, the meaning of the body-fixed blend is inseparable from the distance vector $\hat{R}$, which Oliver has not correctly defined or projected on $\hat{x}$ yet.

We see that Oliver has a strong algebraic background because he is quite good with formal mathematics. Oliver also acknowledged the different meanings that could be designated to positive and negative signs. At first, he tried to collect all the signs with different meanings into his math and simplify with formal mathematics. He spent much of his time reading out the directionality meanings of the remaining signs, flexibly switching among them, comparing and contrasting with the help of the diagram in order to assign appropriate meanings for the signs. Oliver could not eventually resolve his struggles with combining meanings. However, he successfully affiliated other meanings (such as the body-fixed blend, destructiveness in comparative blends) with the comparison between the component field and the $\hat{x}$ or space-fixed blend, which lead him to correct answers. 

Altogether, Oliver's solution path through this problem is:
\begin{align}
&(\Circle, \blacksquare, \triangle) \rightarrow
\Square \rightarrow    
\Circle \rightarrow  
\Square \rightarrow
\Circle \rightarrow
\label{eqn:oliver}\\
\nonumber
&\Square \rightarrow
(\Square, \triangle)
\rightarrow
(\Circle, \Square) \rightarrow
(\Circle, \Square)
\end{align}
In contrast to each example solutions (Equations \ref{eqn:exemplar1}-\ref{eqn:exemplar4}), Oliver's solution is considerably longer and uses all of the blends between \textit{direction} and \textit{sign}.

\section{Charlie Case Study}

In contrast to the case of Oliver, we will look at Charlie's reasoning with signs in the process of solving the same problem of interest. Charlie also struggled with the ways in which signs in the numerator indicate directionality. Apart from that, he had particular difficulty determining distances between point charges and field points. He is good at metacognitively communicating with the instructor but lacked stability and consistency in his reasoning throughout solving the problem. 

Charlie first recorded the electric field direction caused by each charge on the diagram - ``Positive charge - out, negative charge - in, [in the middle] it's comming over to the negative.'' ( Fig. \ref{fig:efields}). He then recorded Coulomb's law with total charge $Q=-2q$ (Table \ref{tab:charlie}, move 1). Charlie then treated $\hat{R}$ and $R$ separately where he indicated $\hat{R}$ as being ``\textit{along} the $x$ direction ($\Circle$)''; hence $\hat{R} = \hat{x}$ and $R=2a$ - ``that's $a$ plus another $a$'' ($\blacktriangle$) (Table \ref{tab:charlie}, move 2). He was quite suspicious of the result - ``I feel like uh\dots it's more for a\dots like a point [charge], right?''. However, Charlie still inserted signs and concluded about the electric field for each region - ``Well, for [region 1], it is in the positive direction, then [in region 2] it is in the negative direction, and [in region 3] it is in the positive direction.'' (Table \ref{tab:charlie}, move 2). Charlie's reasoning was vague here, but we argue that he was using the space-fixed ($\Square$) with helps of the total vector fields on the diagram he just drew. Realizing that the electric field varies along the $x$ axis, Charlie decided to consider the electric field in different regions using the superposition theorem (Fig. \ref{fig:efields}).

\begin{table*}[tb]
\begin{tabular}{|c|p{2cm}|c|l|p{6cm}|}
\hline
Move & Region  & Blend      & Reason                             & Math                 \\
\hline
1	&	General	& $\Circle$ & $\hat{R}$ and $\hat{x}$ parallel  & $\hat{R}$ = $\hat{x}$   \\
    &   & $\blacktriangle$ & $x_{-q}$ and $x_{+q}$ are unsigned number & $R=a+a=2a$ \\
    &   & $\triangle$ & Coulomb's Law  & $\vec{E}=k\frac{(-2q)}{R^2}\hat{x}$ \\
\hline
2   &   1 & $\Square$ & $\hat{E}_{1}$ points right & $\vec{E}_1=k\frac{2q}{R^2}\hat{x}$    \\
    &   2   & $\Square$ & $\hat{E}_{2}$ points left & $\vec{E}_2=k\frac{-2q}{R^2}\hat{x}$ \\
    &   3   &  $\Square$ & $\hat{E}_{3}$ points right & $\vec{E}_3=k\frac{2q}{R^2}\hat{x}$ \\
\hline
\hline
3	&	1 	&	$\blacktriangle$ & $x$, $x_{+q}$, $x_{-q}$ are unsigned numbers & $R_{+q}=x+a$, $R_{-q} = x-a$    \\
    &       &   $\triangle$ & Coulomb's Law & $\vec{E}_1=\vec{E}_{+q}+\vec{E}_{-q}=k{[}\frac{q}{(x+a)^2}+\frac{-q} {(x-a)^2}{]}$   \\
\hline
4	&   1 	&	$\Circle$ & $\hat{E}_{+q}$ and $\hat{E}_{-q}$ distructive & $\vec{E}_1=\vec{E}_{+q}-\vec{E}_{-q}=k{[}\frac{q}{(x+a)^2}+\frac{q}{(x-a)^2}{]}\hat{x}$ \\
\hline
5 	& 	1	&	$\Square$   & $\hat{E}_{+q}$ points left, $\hat{E}_{-q}$ points right & $\vec{E}_1=\vec{E}_{+q}+\vec{E}_{-q}=k{[}\frac{-q}{(x+a)^2}+\frac{q}{(x-a)^2}{]}\hat{x}$ \\
 \hline
 \hline
6	&	2 	&	$\blacksquare$ & $x_{-q}$ is a signed number &  \\
    &       &   $\blacktriangle$ & summing distance & $R_{-q}=x+(-a)$   \\
\hline
\hline
7	&	1 	&	$\blacksquare$ & $x_{-q}$ is a signed number &  \\
    &       &   $\blacktriangle$ & $x_{+q}$ and $x$ are unsigned numbers, subtracting distance & $R_{-q}=-x-(-a)$, $R_{+q}=-x-a$   \\
\hline
8 	& 	1	&	$\blacksquare$ & checking tool & $R_{-q}=x+a$, $R_{+q}=x-a$\newline $\vec{E}_1=\vec{E}_{+q}+\vec{E}_{-q}=k{[}\frac{-q}{(x-a)^2}+\frac{q}{(x+a)^2}{]}\hat{x}$ \\
 \hline
 \hline
9 & 2 & $\blacksquare$ & $x_{-q}$ and x are unsigned numbers & $R_{-q}=x-(-a) = x+a$  \\
    &   & $\blacktriangle$ & $x_{+q}$ and x are signed numbers & $R_{+q}=-x+a$ \\
    &   & $\triangle$ & Coulomb's Law & $\vec{E}_2=\vec{E}_{-q}+\vec{E}_{+q}=k{[}\frac{-q}{(x+a)^2}+\frac{q}{(x-a)^2}{]}\hat{x}$ \\
\hline
10 & 2 & $\Square$ & $\hat{E}_{+q}$ and $\hat{E}_{-q}$ points left & $\vec{E}_2=\vec{E}_{-q}+\vec{E}_{+q}=k{[}\frac{-q}{(x+a)^2}+\frac{-q}{(x-a)^2}{]}\hat{x}$ \\
\hline
\hline
11 & 3 &    &   & $\vec{E}_3=\vec{E}_{-q}+\vec{E}_{+q}=k{[}\frac{-q}{(x+a)^2}+\frac{q}{(x-a)^2}{]}\hat{x}$ \\
\hline
\end{tabular}
\caption{Charlie derives solution. Space-fixed blend = $\square$. Body-fixed blend =$\triangle$.  Comparative blend =$\Circle$. Signed blend = $\blacksquare$. Unsigned blend =$\blacktriangle$.  \label{tab:charlie}}
\end{table*}

\begin{figure*}
  \includegraphics[width=\linewidth]{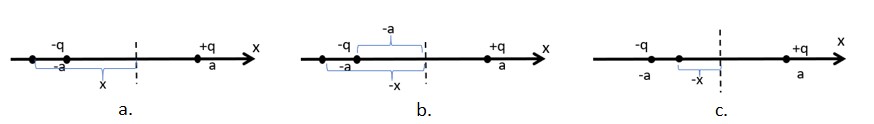}
  \caption{Charlie sketches on the diagram.}
  \label{fig:sketch}
\end{figure*}

Starting with region 1, Charlie spent some time to figure out what is the distance $R$ and how to define $x$ (Fig. \ref{fig:sketch}a). He obtained $R_{+q}=x+a$ and $R_{-q}=x-a$, which indicates his use of unsigned number ($\blacktriangle$) sense and blended the effect of the charges' signs ($\triangle$) into the field expression (Table \ref{tab:charlie}, move 3). With unsigned $x$, Charlie realized that the total field, which must be in the positive $\hat{x}$ direction as he previously mentioned (Table \ref{tab:charlie}, move 2), now was negative because ``the field [$|\vec{E}_{-q}|$] is going to be bigger than [$|\vec{E}_{+q}|$]''. To make the total field positive, Charlie thought that he should have a negative sign connecting the two terms - ``[\dots] but I don't know why, I think I am missing something that makes my negative sign''. 

Later, he inserted a negative sign between two terms and ascribed it to \textit{destructiveness} ($\Circle$) of the fields (Table \ref{tab:charlie}, move 4) as ``[$\vec{E}_{+q}$] points [leftwards] [\dots] [$\vec{E}_{-q}$] points [rightwards] [\dots] Contribution \textit{subtracts}''. However, the result then had both component fields in the positive $\hat{x}$ direction. He then added a negative sign in front of the $+q$ charge to account for its contribution of ``\textit{pushing} away'' in the $-\hat{x}$ direction. His pointing gestures to the left and to the right suggested a space-fixed blend ($\Square$). Charlie reached a his ending solution and checked the sum to make sure the total field is positive ``because [$\vec{E}_{+q}$] has negative direction, now [$\vec{E}_{-q}$] will be bigger [\dots ] so [$\vec{E}_1$] would be positive'' (Table \ref{tab:charlie}, move 5). Notably, even though the result is correct with positive $x$, later when treating $x$ as a negative signed number, Charlie had to come back to this solution and make changes in these terms accordingly.

Moving on to the middle region, Charlie soon realized that $x$ can be negative to the left of the origin and that he had been treating $x_{+q}$, $x_{-q}$, and $x$ as unsigned numbers in region 1. From this point forward, Charlie encountered an array of difficulties and inconsistencies when mixing unsigned and signed blends in his attempts to find distance. In his first try for a field point to the right of the origin, Charlie clearly stated ``[$R_{-q}$] is gotta be $x$ plus the negative $-a$\dots So this \textit{distance} right here is \textit{distance} $x$ and [$x_{-q}$] is like a \textit{distance} $-a$. So this \textit{distance} [$R_{-q}$] should be $x-a$''. This means, he treated $x_{-q}$ as a signed number ($\blacksquare$) but summed ($\blacktriangle$) it with distance $x$ to get $R_{-q}=x-a$ (Table \ref{tab:charlie}, move 6). Following similar failed attempts, Charlie even suggested switching the direction around, but he quickly realized that would not solve his problems. Therefore, he continued with the given axis and decided to extract the signs out of the variable $x$ and make it stay positive; Charlie preferred working with unsigned numbers. 

Charlie then went back to his solution in region 1 and consistently made changes in his mathematical expression. His marks on the diagram (Fig. \ref{fig:sketch}b) and gestures showed that, even when he treated $x_{-q}=-a$ as a signed number ($\blacktriangle$) and $x$ as an unsigned number ($\blacksquare$), he subtracted these two distances ($\blacksquare$) - ``\textit{negative} $x$ \textit{minus negative} $a$'' - to obtain $R_{-q}=-x-(-a)=-x+a$ (Table \ref{tab:charlie}, move 7). For $R_{+q}=-x-a$, Charlie's reasoning was vaguer. Charlie seemed to have his sense of signed and unsigned numbers extremely entangled.

To help Charlie become unstuck, the instructor introduced a checking tool that can help activate knowledge of signed numbers and the signed blend ($\blacksquare$). For instance, when the field point is put right on top of $-q$, Charlie found that the value of $x=x_{-q}=-a$ and the distance $R_{-q}$ should equal zero. From his current expression of $R_{-q}=-x+a$, Charlie decided to bring the negative sign back into $x$ - ``Can we just keep it like this and then the negative sign [\dots ] will be \textit{inside} the $x$ whenever it is on \textit{the other side} of the axis?'' - and treated $x$ as a signed number ($\blacksquare$), turning $R_{-q}$ into $R_{-q}=x+a$. Similarly, Charlie obtained $R_{+q}=x-a$ and accordingly replaced the denominators in the field $\vec{E}_1$ (Table \ref{tab:charlie}, move 8).

Despite getting the correct answer for region 1, Charlie was still inflexible with the checking tool, and especially tended to treat $x$ as an unsigned number. When starting over in region 2, Charlie marked a field point to the left of the origin and recorded the distance as $-x$ on the diagram (Fig. \ref{fig:sketch}c). He then paused for a while, realizing his unsigned approach: ``See if I don't put the negative [in front of $x$ in his recent answer], why am I now calling this \textit{negative} $x$ and even earlier called that \textit{negative} $x$?'' Charlie then went back to the signed blend, which lead him to $R_{-q}=x-(-a)=x+a$: ``See, what my thought is,\dots [$R_{-q}$] should be $x$ \textit{minus negative} $a$. Okay, that makes sense [\dots] Because then, if $x$ is negative, as it approaches [$-a$], it will get to $0$.'' He then quickly arrived at $R_{+q}=-x+a$ without much additional reasoning. This was because when considering a positive number, for instance $x_{+q}$, there is more flexibility in choosing signed and unsigned blends. Hence, the reasoning is more straightforward.

In this case, his diagram and verbal reasoning of ``\textit{negative} $x$ \textit{plus} $a$'' suggested that he might have used the blend of unsigned number to sum ($\blacktriangle$) the two distances $a$ and $(-x)$, which is positive now. Plugging in $R_{-q}$ and $R_{+q}$ and blending the signs of the charges ($\triangle$) into the superposition theorem, Charlie arrived at a field expression that disagreed with the relative directions among $\vec{E}_{-q}$, $\vec{E}_{+q}$, and $\hat{x}$ (Table \ref{tab:charlie}, move 9). Adjusting the sign of each term such that ``[the fields] are both \textit{pushing} in the $-\hat{x}$ direction [pointing \textit{left}] ($\Square$)'', Charlie arrived at the correct expression for the electric field in region 2 (Table \ref{tab:charlie}, move 10).

In region 3, Charlie quickly recorded the total field with no further struggles or explanations (Table \ref{tab:charlie}, move 11). We see that Charlie had successfully affiliated the meaning of signs in denominators with the relative direction between the electric field and $\hat{x}$ after working on the last two regions. Moreover, Charlie's ease in this region could be partially because in this region, signed and unsigned numbers do not matter and hence effective blends of them are not important. 

Charlie could do formal math quickly, and sometimes he skipped a few math steps as well, indicating a facility and ease with algebra. Charlie was also able to connect the destructiveness meaning into the relative direction of two vectors:  ``It should not be \textit{subtracted} from the other, should it? I feel like the electric fields both \textit{add} up together, and in one case, just in this case, it's \textit{negative}''. Apart from the similar procedure of determining the appropriate directionality meaning for signs, however, Charlie also spent a lot of time and effort working with the denominators, where he struggled to employ consistent and helpful ways to think about the distance between two points. The problem is subtly challenging because both variables $a$ and $q$ were introduced as unsigned numbers. Therefore, it seems natural to treat $x$ as an unsigned number as well. In fact, we see via Charlie's diagram and gesture, that he preferred treating $x$ as an unsigned number, where increasing and decreasing the distance was associated with positive and negative signs. Even though Charlie finally made sense of the checking tool and the signed blend, he tended to go back to an unsigned number and unsigned blend whenever he was in a positive region.

Altogether, Charlies's solution path through this problem is:
\begin{align}
&(\Circle, \blacktriangle, \triangle) \rightarrow  
(\Square, \Square, \Square) \rightarrow  
(\blacktriangle, \triangle) \rightarrow
\Circle \rightarrow
\Square \rightarrow 
\label{eqn:charlie}\\
\nonumber
&(\blacksquare, \blacktriangle) \rightarrow
(\blacksquare, \blacktriangle) \rightarrow
\blacksquare \rightarrow
(\blacksquare, \blacktriangle, \triangle) \rightarrow
\Square
\end{align}
In contrast to each example solution (Equations \ref{eqn:exemplar1}-\ref{eqn:exemplar4}), Charlie's solution is considerably longer and uses all of the blends between \textit{direction} and \textit{sign} and \textit{location} and \textit{sign}.  

\section{Discussion: Oliver and Charlie in Comparison}

Oliver and Charlie had similar approaches to their solution derivations, especially in the way they sought signs with appropriate meanings to include in their mathematical expressions. Their verbal, diagrammatic, gestural, and mathematical reasoning indicated that they could distinguish different meanings associated with the signs. In addition, both students showed their sound background in algebra by efficiently carrying out operations involving negative signs such as adding, subtracting, and multiplying. These formal mathematical steps assisted them in thinking about the signs with different meanings and affiliating those meanings together. Therefore, their struggles with the problem were not because they had not been equipped with the necessary mathematical tools but could be plausibly attributed to the challenging task of understanding and manipulating the physical meaning embodied in algebraic signs.

Charlie clearly struggled with the problem much more than Oliver because he struggled to comprehend not only the directionality meaning, but also the location meaning of the signs - how to treat coordinates and what blends to use to find the distance between two points. Oliver seemed to have a better physical sense of the situation from recalling Coulomb's law and the superposition theorem whereas Charlie inappropriately treated the system as just a bigger point charge initially. However, there still exists a pattern in their directionality reasoning. Starting considering directionality, Both Charlie and Oliver naturally collected the signs that embedded the charge effect of the body-fixed coordinates ($\triangle$) and the relative relation $\hat{R} = \hat{x}$ ($\Circle$) into the field expression. In the second region, they repeated the choice of the body-fixed coordinates ($\triangle$). Both students tried hard to combine this meaning ($\triangle$) with others and eventually affiliate it with the signs showing the directions of the vectors $\vec{E}_{+q}$ and $\vec{E}_{-q}$, either in the directional comparison with each other ($\Circle$) or in their leftwards - rightwards direction ($\Square$), with the latter tends to be more favored.

Remarkably, in their first attempt at fitting their expression for region 1 with the diagram, both students considered destructiveness as being separate from opposite directions. Even though Charlie thought that both destructiveness and opposite direction are embodied in the same signs, he eventually inserted an extra negative sign to account for a destructive combination. Later, we observed that interference no longer bothered Charlie or Oliver. In region 2, this could have been because the effect of constructive interference is consistent with the positive signs in the superposition theorem. For region 3, this could have been because both Charlie and Oliver had already firmly affiliated the interference effect with signs that indicate relative direction.

Oliver's solution path: (Equation \ref{eqn:oliver})
\begin{align*}
&(\Circle, \blacksquare, \triangle) \rightarrow
\Square \rightarrow    
\Circle \rightarrow  
\Square \rightarrow
\Circle \rightarrow
\\
&\Square \rightarrow
(\Square, \triangle)\rightarrow
(\Circle, \Square) \rightarrow
(\Circle, \Square)
\end{align*}

Charlie's solution path: (Equation \ref{eqn:charlie})
\begin{align*}
&(\Circle, \blacktriangle, \triangle) \rightarrow  
(\Square, \Square, \Square) \rightarrow  
(\blacktriangle, \triangle) \rightarrow
\Circle \rightarrow
\Square \rightarrow 
\nonumber \\
\nonumber
&(\blacksquare, \blacktriangle) \rightarrow
(\blacksquare, \blacktriangle) \rightarrow
\blacksquare \rightarrow
(\blacksquare, \blacktriangle, \triangle) \rightarrow
\Square\\
\end{align*}

Notably, Oliver and Charlie might have obtained correct answers sooner if they could have appropriately defined the relative direction between the distance vector $\hat{R}$ and $\hat{x}$. Both correctly defined the magnitudes of $R_{+q}$ and $R_{-q}$ but failed to determine the directions of $\hat{R}_{+q}$ and $\hat{R}_{-q}$ as they tended  to treat magnitude and direction separately. Consequently, neither student could reason consistently between the signs in $\pm q$ and $\hat{R}$ that occur in Coulomb's law, and the signs that resulted from the blends of relative direction between $\vec{E}_{+q}$, $\vec{E}_{-q}$, and $\hat{x}$. 

\section{Conclusion}
In the context of an electrostatics problem, we built five different blends to illustrate how different meanings of positive and negative signs might be constructed. In the blended space of directionality with sign, different elements projected into the blended space could lead the signs to be associated with space-fixed coordinates, body-fixed coordinates, or comparative directions. Similarly, in the blended space of location with sign, the signs could account for location on the left or right side of the origin, as well as with operations for determining distance between two points on the axis according to combinations of signed and unsigned numbers.  This is important because prior work in PER suggests that each pair of spaces has only one resultant blend; the difficult work is in selecting input spaces, not selecting the blend.  

Among the six blends, the blends between location and sign could closely relate to the triple natures of the minus signs in elementary algebra \cite{vlassis2004making}. For instance, the negative sign in the negative signed number $x=-a$ purely carries the unary function (a location relative to an origin). With the sense of getting shorter or longer distances in an unsigned blend, we claim that the negative sign in such expressions as $(x-a)$ is used as a binary function of taking away. Especially when treating $x$ as a positive number and realizing that $x$ could be negative in the left region, Charlie's attempts to extract the negativity out of the variable $x$, turning it from a singed to an unsigned number, suggest the minus sign in $-x$ falls into the symmetry category on Vlassis's map. On rare occurrences, we might see all three meanings of the negative sign appear in a single mathematical expression: $R_{-q}= -x-(-a)$ (Table 4, move 7). 

Although helpful, the map developed by Vlassis does not fully account for the negative signs that come from vector algebra. The negative sign associated with opposite relative direction could be symmetrical (in the case of a relative sign) or binary (in case of the connecting sign in a mathematical expression of interference). In addition, there seems to be no good match for the body-fixed coordinate or the space-fixed coordinate convention. Since the map of different uses of the negative sign is highly context-sensitive, the transposition of meaning from elementary algebra to physics will require inevitable adjustment and expansion. 

By considering five different blends within the context of conceptual blending theory, we analyzed two case studies to see why an introductory-level physics problem turns out to be challenging to upper-level undergraduate students. We scrutinized the existence of these blends throughout the students' reasoning. Most of the blends were constructed and employed subconsciously and without effort, while others were employed deliberately and with more effort. Students were observed to continuously switch among the meanings to use, combine those meanings, and affiliate meanings to the last signs left in the expression. We claim that the problem is challenging for students because there are not only multiple signs coming within a single expression, but also multiple meanings with which the signs could associate, where students need to carefully choose what blends to use.  This is important because blending work outside of PER suggests that the process of blending is fast and automatic, whereas work within PER indicates that it can be slow and difficult. 

Adding to the current literature of multiple representations associated with a concept, we affirm that the multiple physical meanings that can be associated with a simple algebraic symbol could also cause difficulties. One possible source of these difficulties is students' lack of understanding of the meaning embedded in a sign. For example, both Oliver and Charlie never explicitly talked about the body-fixed meaning but kept using them unwittingly as having been drawn from an authoritative source. Eventually, Charlie affiliated the body-fixed blend with other blends, while Oliver decided to just ignore it. An implication from this study is that students need more help to better understand and be able to distinguish the deeper meaning of algebraic signs. This might then lead to more successful problem solving. 


\section{Acknowledgements}
This material is based upon work supported by the National Science Foundation under Grant DUE-1430967.  An earlier version of a subset of the Oliver data was presented at ICLS 2018.  We are deeply indebted to our beta-readers, Julie 
Nurnberger-Haag, the PEER-Rochester thesis writers group, and the KSUPER group, and to encouragement from Suzanne Brahmia.  Jeremy Smith copyedited this paper.  Any remaining issues are solely our own. 

\bibliographystyle{plainnat}

\end{document}